\documentclass[aps,showpacs]{revtex4}
\usepackage{graphicx}
\def\no{\noindent}
\def\bc{\begin{center}}
\def\ec{\end{center}}

\def\beq{\begin{equation}}
\def\eeq{\end{equation}}

\def\d{\downarrow}
\def\u{\uparrow}

\def\bj{{\bf j}}

\begin{document}


\title{Dynamics of finite Fermi-Hubbard and Bose-Hubbard systems}

\author{K. Ziegler}
\affiliation{Institut f\"ur Physik, Universit\"at Augsburg\\
D-86135 Augsburg, Germany}
\date{\today}
\begin{abstract}
This paper analyzes dynamical properties of small Fermi-Hubbard and Bose-Hubbard systems,
focusing on the structure of the underlying Hilbert space. We evaluate time-dependent
quantities such as the return probability to the initial state and the spin imbalance 
of spin-1/2 fermions. The results are compared with recent experimental observations in ultracold gases. 
For the symmetric two-site Fermi-Hubbard model we find that the spin imbalance and the return 
probability are controlled by two and three frequencies, respectively.
The spin imbalance and the return probability are identical for the asymmetric Falicov-Kimball limit
and controlled by only one frequency. In general, the transition probabilities between the
initial state and energy eigenstates depend strongly on the particle-particle
interaction. This is discussed for ``self trapping'' of spinless bosons in 
a double-well potential. We observe that the available Hilbert space is reduced
significantly by strong interaction.
\end{abstract}


\maketitle

\section{Introduction}

The dynamics of many-body quantum states has been studied with high
accuracy in ultracold gases, either for bosons 
\cite{foelling07,spielman07,salger07,trotzky08,wuertz09} or for
fermions \cite{shin06,fermischool07}. The main advantage of employing an ultracold gas 
is that there are many parameters in the experiment that can be adjusted to control the initial state
as well as the dynamics of the system. This means in
particular that an ultracold gas can be prepared in almost any state 
$|\Psi_0\rangle$, not only in the ground state. After its preparation at time $t=0$, 
the state evolves in time $t>0$ for a system with Hamiltonian $H$,
which describes  the kinematics and the interaction of the atomic particles, 
according to 
\beq
|\Psi_t\rangle = e^{-iHt}|\Psi_0\rangle  \  .
\label{evolution0}
\eeq
In the following we will consider the simplest possible case, namely a model with
two sites. As a first example we study two fermions with spin 1/2, a second 
example is a system of $N$ spinless bosons in a double well. Despite of their simplicity, these examples reveal
dynamical properties that may have also implications to more complex systems.
For instance, an interesting question is whether or not all states of the underlying
Hilbert space can be reached from a given initial state, which time scales appear
and how this can be controlled by adjusting the parameters of the system such as
the tunneling rate and the interaction strength. 

The dynamics of atomic systems with a finite number of atoms has been an active
field of research for several decades, with a focus on oscillating behavior between
different atomic states \cite{cohen,lewenstein07}. More recently, ultracold gases in an optical
lattice have been a successful platform for such investigations 
\cite{foelling07,esteve08,trotzky08}. In particular, small systems of spin 1/2
fermions have also attracted considerable attention by the quantum computation 
community \cite{lewenstein07,cirac07}. 
A typical initial state is a (paramagnetic) Mott state, where the orientation of the
individual spins can be used for information storage. Since for $N$ particles this state 
has an exponential degeneracy $2^N$, the time evolution in terms of, for instance, 
a Hubbard Hamiltonian can lead to a complex dynamics. Of particular interest is how the
spin population of a given site changes with time.

Experiments with a Bose gas in an optical double well have revealed that
the population dynamics is controlled by direct tunneling of atoms and by a second-order
(superexchange) process, characterized by two frequencies  \cite{foelling07,esteve08}.
The appearance of the characteristic frequencies in finite bosonic systems
was calculated in strong-coupling perturbation theory \cite{duan03,rey07}
and in time-dependent Hartree-Fock approximation \cite{milburn97,zoellner08}.
In a more recent paper by Trotzky et al. \cite{trotzky08} the spin imbalance of two spin-1/2
atoms in a double well was studied. Such a system can be described by a two-site
Bose-Hubbard model, which is often called the two-mode approximation \cite{milburn97}.
With increasing values of $J/U$, where $J$ is the tunneling rate and $U$ 
is the local interaction strength,  the oscillating spin imbalance was 
increasingly influenced by a second frequency. This implies that both, the tunneling rate
of the atoms as well as the interaction strength, determine the dynamics of the
many-body system. In this paper we will study this effect in terms of a Fermi-Hubbard model.
The focus will be on the return probability of the many-body state. This
is an important quantity for recovering information which was stored in the initial
state.    

There are several options for an analytic calculation of physical quantities in 
a finite many-body system.
Perturbative methods are very successful and reliable approaches to physical
problems and can be considered as exact if their regimes of validity are respected.
On the other hand, they fail in most cases near a singularity, where their
validity is violated. Nevertheless, they can still be used in these cases as 
approximation methods by using an asymptotic partial summation of the 
perturbation series. Hartree-Fock approximations belong to this type of approximations 
and have been used in the case of finite many-body systems \cite{milburn97,zoellner08}.
Other self-consistent approaches to many-body systems have been very successful,
such as the dynamical mean-field theory \cite{georges07}. 
A disadvantage of all these approaches is that the (nonlinear) self-consistent equations 
are very complex, such that their treatment usually requires intensive numerical work.

An alternative to the perturbative approach and to self-consistent approximations
is the recursive projection method (RPM) \cite{feshbach67,ziegler}. It is 
a systematic exploration of the Hilbert space, using a recursive calculation of
the resolvent $(z-H)^{-1}$. The latter can be obtained from the Laplace transformation 
of Eq. (\ref{evolution0}) (cf. Sect. \ref{resolvent01}). 
The RPM enables us to extract systematically the
poles of the resolvent in a subspace of the underlying physical Hilbert space. 
This method, combined with a truncation of recursion, is related to the Lanczos 
procedure \cite{lanczos}.
The RPM has been explained elsewhere in the literature \cite{feshbach67,ziegler}, 
here we give only a brief summary and apply it to the Hubbard models 
in Sects. \ref{dyn1}, \ref{dyn2}.

The paper is organized as follows: 
After a short description of the RPM in Sect. \ref{rpm01}
we define the Fermi-Hubbard and the Bose-Hubbard model, 
the many-body return probability, the spin imbalance and the
many-body spectral density in Sect. \ref{model}.
In Sect. \ref{resolvent01} the dynamics is discussed in terms of the resolvent.
Then the recursive projection method is applied to the two-site Fermi-Hubbard
(Sect. \ref{dyn1}) 
and to the two-site Bose-Hubbard model (Sect. \ref{dyn2}) to derive 
dynamical properties of these systems.  Finally, the results
of the RPM are discussed in detail in Sect. \ref{results}.
In particular, frequencies and transition probabilities are calculated and their 
dependence on the model parameters are discussed.

\section{Recursive Projection Method}
\label{rpm01}

The structure of our physical system is completely determined by a Hamiltonian $H$ 
that acts on Hilbert space ${\cal H}$. Then the central idea of the RPM is 
that the dynamics starts from an initial state which lives in a subspace 
${\cal H}_0\subset {\cal H}$. It should consist of a basis that is 
dynamically separable, meaning that the Hamiltonian does no allow to 
move directly from one basis state to another. 
Once this subspace has been chosen specifically, 
the rest of the RPM is entirely determined by the Hamiltonian $H$, and
the dynamics, given by the time-evolution operator, decides which part of ${\cal H}$
is relevant. This depends on the energies associated with the remaining
Hilbert space and on the transition probabilities. 
Now we project with projector $P_0$ onto the Hilbert
space ${\cal H}_0$. The corresponding projected resolvent is $G_0(z)=P_0(z-H)^{-1}P_0$.
Then the RPM includes two steps:

\no
(1) Hilbert space ${\cal H}_{2j+2}$ ($j=0,1,...$) is created by acting with 
the operator $({\bf 1}-P_0-P_2-\cdots -P_{2j})HP_{2j}$ on ${\cal H}$. In other
words, a basis set from all the states created by 
$({\bf 1}-P_0-P_2-\cdots -P_{2j})HP_{2j}$ is a basis of ${\cal H}_{2j+2}$.

\begin{figure}
\begin{center}
\includegraphics[width=3cm,height=3cm]{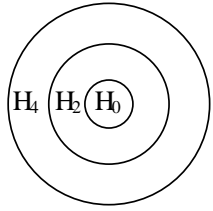}
\caption{
The schematic structure of the sequence of Hilbert spaces ${\cal H}_{2j}$ for $j=0,1,...$, 
created from the initial Hilbert space ${\cal H}_0$ by applying the Hamiltonian $H$, 
resembles a Russian doll. The resolvent $G_0$ is projected onto ${\cal H}_0$ and
the outer shells contribute as poles to $G_0$ (cf. Sect. \ref{rpm01}). The recursion
removes one shell (mother doll) after the other until one reaches the baby doll ${\cal H}_0$. 
}
\label{russiandoll}
\end{center}
\end{figure}

\no
(2) evaluating the resolvent $G_{2j}$ on ${\cal H}_{2j}$ through the recurrence relation
\beq
G_{2j}=\left(z-H'_{2j}\right)_{2j}^{-1}
\label{projected3}
\eeq  
with the effective Hamiltonian $H'_{2j}$ on ${\cal H}_{2j}$:
\beq
H'_{2j}=P_{2j}HP_{2j}+P_{2j}HG_{2j+2}HP_{2j} \ .
\label{effham}
\eeq
{\it remarks:}
(I) It should be noticed that the construction of the sequence ${\cal H}_{2j}$ 
($j=1,2,...$) implies that Hilbert space ${\cal H}_{2j}$ is orthogonal
to ${\cal H}_{2j'}$ for $j'\ne j$. Moreover, the recurrence relation never
returns to previously visited subspaces. This can be represented schematically as 
a Russian-doll structure, shown in Fig. \ref{russiandoll}.
(II) The recursion terminates in a finite dimensional Hilbert space with 
the effective Hamiltonian
\[
H'_{2n}=P_{2n}HP_{2n} \ .
\]
This is the only effective Hamiltonian that is explicitly given, provided we
know the projection $P_n$. In order to use this as the initial effective Hamiltonian,
we introduce $k=n-j$ as the running index in the recurrence relation. Then we have
\beq
G_{2(n-k)}=\left(z-H'_{2(n-k)}\right)_{2(n-k)}^{-1}
\ ,
\label{projected4}
\eeq  
and with $g_k\equiv G_{2(n-k)}$, $h_k\equiv H'_{2(n-k)}$ we obtain the recurrence
relation
\beq
g_k=\left(z-h_k\right)_{2(n-k)}^{-1}
\ ,
\label{projected4a}
\eeq  
where
\beq
h_k=P_{2(n-k)}HP_{2(n-k)}+P_{2(n-k)}Hg_{k-1}HP_{2(n-k)}, \ \ h_0=P_{2n}HP_{2n}
\ .
\label{effham1}
\eeq
This means in terms of the Russian doll that, in order to evaluate $G_0$ 
(i.e. the baby doll), we must inherit
the properties of all generations of mother dolls $G_{2j}$ ($j=1,...,n$), 
using th recurrence relation iteratively.

\section{Model}
\label{model}

The  Fermi-Hubbard (FH) model describes locally interacting fermions with spin $\sigma=\u ,\d$,
the Bose-Hubbard (BH) model locally interacting spinless bosons. 
It is defined by the Hamiltonian $H=H_J+H_I$,
where $H_J$ is the tunneling term between the sites $1$ and $2$. For fermions this reads
\beq
H=-\sum_{\sigma=\u,\d}J_\sigma (c_{1\sigma}^\dagger c_{2\sigma}+c_{2\sigma}^\dagger c_{1\sigma})
+U(n_{1\u}n_{1\d} +n_{2\u}n_{2\d})
\label{hamferm}
\eeq
with fermionic creation operators $c_{\bj\sigma}^\dagger$, annihilation operators $c_{\bj\sigma}$,
and density operators $n_{\bj\sigma}= c_{\bj\sigma}^\dagger c_{\bj\sigma}$. There are two
special cases, the symmetric FH model with $J_\d=J_\u$ and the Falicov-Kimball limit
$J_\d>0$, $J_\u=0$ \cite{falicov69}. 

For bosons with bosonic creation operators $b_{\bj}^\dagger$, annihilation 
operators $b_{\bj}$, and density operators $n_{\bj}= b_{\bj}^\dagger b_{\bj}$
the Hamiltonian reads
\beq
H=-J(b_{1\sigma}^\dagger b_{2\sigma}+b_{2\sigma}^\dagger b_{1\sigma})
+U(n_{1}^2+n_{2}^2)
\ .
\label{hambos}
\eeq
This Hamiltonian is also
known as the two-mode approximation of a continuous Bose gas
in a double-well potential \cite{milburn97}. 

For fermions the local (Hubbard) interaction $H_I=U(n_{1\u}n_{1\d} +n_{2\u}n_{2\d})$
can be diagonalized with product (Fock) states 
\beq
|\sigma_1,\sigma_2 \rangle \  \  \
(\sigma_\bj=0, \u,\d, \u\d )  \ ,
\label{prodstate}
\eeq
where this product state is a paramagnetic Mott state in the restricted case 
$\sigma_\bj=\u,\d$ (singly occupied sites). Any eigenvalue of $H_I$ with respect
to a Mott state is zero. 

The interaction term of the BH model reads
$
H_I=U(n_1^2+n_2^2)
$, whose eigenstates are also product states $|n_1,n_2\rangle$ ($n_j\ge 0$)
with eigenvalues $U(n_1^2+n_2^2)$. The tunneling term gives for these states
\[
H_J|n_1,n_2\rangle
=-J(b_2^\dagger b_1+b_1^\dagger b_2)|n_1,n_2\rangle
\]
\[
=-J\sqrt{(n_2+1)n_1}|n_1-1,n_2+1\rangle-J\sqrt{n_2(n_1+1)}|n_1+1,n_2-1\rangle
\ .
\]
States with $n_\bj-1<0$ do not exist.

In general, the eigenstates $|E_j\rangle$ of a Hamiltonian $H$ with energy $E_j$
\[
H|E_j\rangle=E_j|E_j\rangle
\]
can be used to describe the time evolution of an initial state $|\Psi_0\rangle$ 
at time $t=0$ to the state $|\Psi_t\rangle$ at later time $t>0$ by
\beq
|\Psi_t\rangle =e^{-iHt}|\Psi_0\rangle =\sum_j e^{-iE_jt}|E_j\rangle\langle E_j|\Psi_0\rangle \ .
\label{spectralrep}
\eeq
There are two interesting quantities directly related to this expression. One is the return 
probability to the initial state $P(t)=|\langle\Psi_0|\Psi_t\rangle |^2$, the other is
the spin imbalance between the two sites in a double well \cite{trotzky08}
\beq
N_{1,2}(t)=\frac{1}{2}
\langle\Psi_t|n_{\u 1}-n_{\d 1}+n_{\d 2}-n_{\u 2}|\Psi_t\rangle \ .
\label{imbalance}
\eeq
The latter describes the exchange dynamics of the two spins $\u$ and $\d$, located at the
two sites. $N_{1,2}$ vanishes if the sites are not singly occupied.

\subsection{The Resolvent}
\label{resolvent01}

A Laplace transformation of $|\Psi_t\rangle$ for positive time gives
the resolvent that acts on the initial state:
\beq
-i\int_0^\infty e^{izt}|\Psi_t\rangle dt
=-i\int_0^\infty e^{izt}e^{-iHt}dt |\Psi_0\rangle
=(z-H)^{-1}|\Psi_0\rangle \ \ \ (Im z>0)
\ .
\label{evol1}
\eeq
Or equivalently, we can express the time evolution operator in terms of the 
resolvent $(z-H)^{-1}$ as
\beq
e^{-iHt}=\int_\Gamma (z-H)^{-1}e^{-izt}{dz\over2\pi i} \ ,
\label{resolvent0}
\eeq
where the closed contour $\Gamma$ encloses all (real) eigenvalues of $H$.
The return probability to the initial state 
is obtained from the inverse Laplace transform of the resolvent through the
relation
\beq
\langle\Psi_0|\Psi_t\rangle=\int_\Gamma e^{-izt}
\langle\Psi_0|(z-H)^{-1}|\Psi_0\rangle {dz\over2\pi i} \ .
\label{integr0}
\eeq
The spectral representation of Eq. (\ref{spectralrep}) allows us to write
\beq
\langle\Psi_0|\Psi_t\rangle=\int_\Gamma e^{-izt}
\sum_j\frac{|\langle E_j|\Psi_0\rangle |^2}{z-E_j}
{dz\over2\pi i}
=\sum_je^{-iE_jt}|\langle E_j|\Psi_0\rangle |^2 \ .
\label{overlap}
\eeq
This means that the matrix element of the resolvent
\beq
\langle\Psi_0|(z-H)^{-1}|\Psi_0\rangle=
\sum_j\frac{|\langle E_j|\Psi_0\rangle |^2}{z-E_j}
\label{expans2}
\eeq
is a rational function with poles $z=E_j$ ($j=0,1,...$).
The many-body spectral density can be calculated from this expression
for $z=E+i\epsilon$ as
\beq
-Im \langle\Psi_0|(E+i\epsilon -H)^{-1}|\Psi_0\rangle
=\sum_j|\langle E_j|\Psi_0\rangle |^2\frac{\epsilon}{\epsilon^2+(E-E_j)^2} \ .
\label{lorenz}
\eeq
By plotting this expression as a function of $E$, we can identify graphically 
the poles $E_j$ ($j=0,1,...$) of $G_0$ and the overlap of $|E_j\rangle$ 
with the initial state. The energy levels $E_j$ are the locations of the Lorentzian peaks and
$|\langle E_j|\Psi_0\rangle |^2/\epsilon$ correspond to the height of the
Lorentzian peaks. 
The knowledge of $E_j$ and $|\langle E_j|\Psi_0\rangle |^2$ enables us to determine
the expression for the dynamical overlap function in Eq. (\ref{overlap}).

\section{Dynamics of two fermions in a double well}
\label{dyn1}

Considering only two fermions with opposite spin, the Hamiltonian acts
on a four-dimensional Hilbert space and can be diagonalized directly
with eigenvalues $0,U,U/2\pm\sqrt{U^2/4+4J^2}$ for the symmetric case $J_\d=J_\u\equiv J$.
Also the RPM is simple because it terminates already for $n=1$, after 
creating a single pair of empty
and doubly occupied sites. The effective Hamiltonians then read  
\beq
H'_2=U,\ \ 
H'_0=\frac{1}{z-U}P_0H^2P_{0} \ .
\label{effham2}
\eeq
$H'_0$ can also be expressed as a Heisenberg spin Hamiltonian \cite{ziegler}:
\beq
H'_0=\sum_{<{\bf j},{\bf j}'>}P_0\left[
a_{\u\d}\left(S_{\bf j}^x S_{{\bf j}'}^x+S_{\bf j}^yS_{{\bf j}'}^y\right)
+a_{\u\u} \left(S_{\bf j}^z S_{{\bf j}'}^z-1/4\right)
\right]P_0 ,
\label{heff}
\eeq
with $z$-dependent spin-spin coupling coefficients 
\beq
a_{\u\u}=2\frac{J_{\uparrow}^2+J_{\downarrow}^2}{U-z} , \ \ 
a_{\u\d}=4\frac{J_{\uparrow}J_{\downarrow}}{U-z} 
\label{coupl1a}
\eeq
and with the spin-1/2 operators 
\[
S^j={1\over2}\pmatrix{
c^\dagger_\u , & c^\dagger_\d \cr
}\cdot\sigma^j\pmatrix{
c_\u \cr
c_\d \cr
}
\hskip0.5cm
(j=x,y,z), 
\]
where $\sigma^j$ ($j=x,y,z$) are Pauli matrices. The spin components
read explicitly
\beq
S^x={1\over2}(c_\u^\dagger c_\d + c_\d^\dagger c_\u) ,
\hskip0.5cm
S^y={-i\over2}(c_\u^\dagger c_\d - c_\d^\dagger c_\u) ,
\hskip0.5cm
S^z={1\over2}(c_\u^\dagger c_\u-c_\d^\dagger c_\d)  \ .
\label{spincomponents}
\eeq
In the symmetric limit $J_{\uparrow}=J_{\downarrow}$ the effective Hamiltonian in 
Eq. (\ref{heff}) becomes an isotropic Heisenberg Hamiltonian.
The singly occupied eigenstates of $H'_0$ are linear combinations of $|\u,\d\rangle$ 
and $|\d,\u\rangle$.
If $\lambda$ is one of the eigenvalues of $(z-U)H'_0$ with
$\lambda=-(J_\u\pm J_\d)^2$, we get for the poles of the projected resolvent $G_0$
\beq
z=\frac{U}{2}\left(1\pm \sqrt{1-4\lambda/U^2}\right)
\sim\cases{
U-\lambda/U \cr
\lambda/U \cr
}
\label{eigenvalue01}
\eeq
from the RPM. The asymptotic expressions hold for $\lambda/U\sim 0$.
For strong interaction parameter $U$, only one pole is accessible by perturbation theory,
such that the appearance of two poles can be understood 
as a simple non-perturbative effect: the Brillouin-Wigner perturbation theory 
in powers of $\lambda/U$ \cite{fradkin} gives only the low-energy pole for $G_0(z)$, 
namely $z=\lambda/U$ \cite{duan03}, and neglects the high-energy pole 
$z\sim U-\lambda/U$. We will see later that the low-energy pole is indeed
negligible for large values of $U$. Experimentally, however, both energies
have been observed in a double well potential \cite{trotzky08}.

\section{Dynamics of $N$ bosons in a double well}
\label{dyn2}

A system of $N$ spinless bosons, distributed over two sites, lives in a
$N+1$-dimensional Hilbert space. Using the basis $|n_1,n_2\rangle$, a
special case for the initial state is $n_1=N$, $n_2=0$: $|\Psi_0\rangle=|N,0\rangle$.
Then all projected spaces ${\cal H}_{2j}$  are one dimensional
and spanned by $|N-j,j\rangle$. This leads recurrence relation (cf. App. \ref{rpmbose})
\beq
g_k=\frac{1}{z-U[k^2+(N-k)^2]-J^2(N-k+1)kg_{k-1}}, \ \ \ g_0=\frac{1}{z-UN^2}
\ .
\label{recurrenceb}
\eeq
Then the projected resolvent for the initial state $|\Psi_0\rangle=|N,0\rangle$ reads 
\[
\langle N,0|G_0|N,0\rangle = \langle N,0|(z-H)^{-1}|N,0\rangle =g_N
\ .
\]
The evaluation of $g_N$ from Eq. (\ref{recurrenceb}) is  a simple task and leads to
a rational function, consisting of a polynomial of order $N+1$ in the denominator.

\section{Results and Discussion}
\label{results}

Using the expressions derived in Sects. \ref{dyn1}, \ref{dyn2} we evaluate the spin imbalance
as well as the return probability of the two-site FH model for symmetric tunneling 
and in the Falicov-Kimball limit. In the second part the dynamics of a two-site BH
model with $N$ bosons is studied by evaluating the many-body return probability and
the many-body spectral density.

\subsection{Two-site Fermi-Hubbard model}
\label{results2shm}

The spin imbalance of Eq. (\ref{imbalance}) can be rewritten in terms of the spin 
operator $S^z$ 
\beq
\langle\Psi_t|S^z_1-S^z_2|\Psi_t\rangle \ ,
\label{si}
\eeq
since $S^z=(n_\u-n_\d)/2$. Then the matrix element for the spin imbalance
of the two-site FH model with respect to the
initial state $|\Psi_0\rangle=|\u,\d\rangle$ becomes after Laplace transformation
(cf. Eq. (\ref{evol1}))
\[
\langle\u,\d|(z-H'_0(z))^{-1}(S^z_1-S^z_2)(z'-H'_0(z'))^{-1}|\u,\d\rangle =
\]
\beq
\frac
{2(U-z)(U-z')(4U^2zz'-4Uzz'^2+2zaU-4z^2z'U+4z^2z'^2-2z^2a
+2z'aU-2z'^2a+a^2-b^2)}
{(4z'^2U^2-8z'^3U+4z'^4+4z'aU-4z'^2a+a^2-b^2)
(4z^2U^2-8z^3U+4z^4+4zaU-4z^2a+a^2-b^2)}
\label{2sh}
\eeq
with $a=2(J_\u^2+J_\d^2)$ and $b=4J_\d J_\u$.
The denominator of this expression is a product of two fourth-order polynomials with 
respect to $z$ and $z'$, which gives four poles for $z$ and for $z'$, respectively:
\[
z_{1/2}=U/2\pm \sqrt{U^2+2a-2b}/2\ , \ \
z_{3/4}=U/2\pm \sqrt{U^2+2a+2b}/2 
\ .
\]
Since $2a\pm 2b=4(J_\u\pm J_\d)^2$, these poles read in terms of the tunneling rates
\beq
z_{1/2}=U/2\pm \sqrt{U^2/4+(J_\u- J_\d)^2}, \ \
z_{3/4}=U/2\pm \sqrt{U^2/4+(J_\u + J_\d)^2} 
\ .
\label{eigenvalues2}
\eeq
The non-interacting limit $U=0$ has the poles $z_{1/2}=\pm(J_\u- J_\d)$ 
and $z_{3/4}=\pm(J_\u +J_\d)$, i.e. they are linear in the tunneling rates.

\begin{figure}
\begin{center}
\includegraphics[width=7cm,height=7cm]{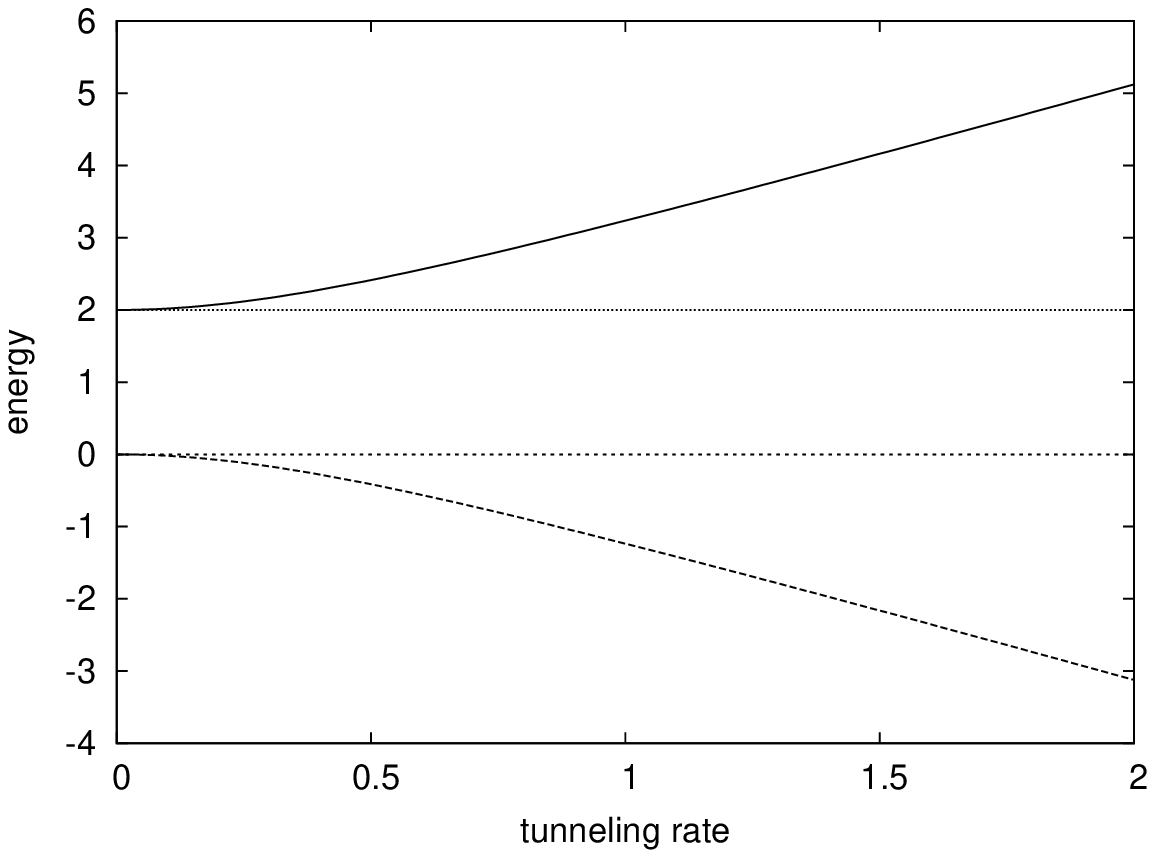}
\includegraphics[width=7cm,height=7cm]{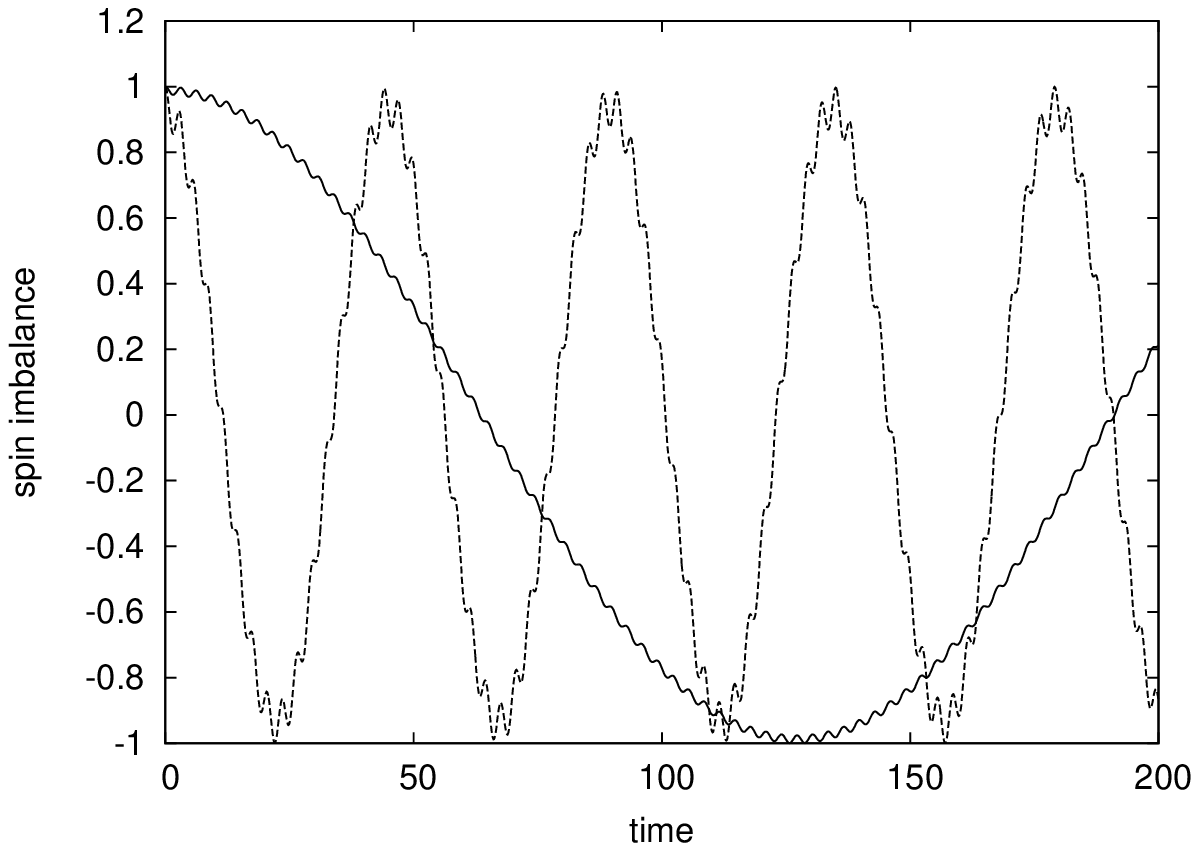}
\caption{
The four energy levels of the symmetric two-site FH model is plotted as a function of the tunneling rate $J$
at $U=2$ (left panel). The double degeneracy at $J=0$ is lifted by the tunneling rate $J$.  
The spin imbalance with initial state $|\Psi_0\rangle=|\u,\d\rangle$ is plotted for $U=2$ (right panel)
with $J/U=0.05$ (full curve), $J/U=0.3$ (dashed curve).
There are two frequencies contributing to each curve, namely $z_{3/4}=U(1\pm \sqrt{1+16J^2/U^2})/2$,
corresponding to the lowest and highest energy level on the left panel.
}
\label{spinimbalance}
\end{center}
\end{figure}

{\it symmetric HF model}:
For simplicity, we first consider a symmetric Hubbard model with
$J_\d=J_\u\equiv J$. Then the matrix element in Eq. (\ref{2sh}) simplifies
because of $a=b=4J^2$, and with $J=1/2$ we get for the spin imbalance the expression
\[
\langle\u,\d|(z-H'_0(z))^{-1}(S^z_1-S^z_2)(z'-H'_0(z'))^{-1}|\u,\d\rangle
\]
\beq
=\frac{2U^2zz'-2Uzz'^2+Uz-2z^2z'U+2z^2z'^2-z^2+z'U-z'^2}
{4(z'U-z'^2+1)z'z(Uz-z^2+1)} 
\ .
\label{spinimb3}
\eeq
It should be noticed here that the poles $z_1=U$ have been canceled by the 
factor $(U-z)(U-z')$ in the numerator of Eq. (\ref{2sh}).
Thus, only the poles $z_2,z_3,z_4$ contribute to the spin imbalance:
\beq
z_2= 0 \ , \ \
z_{3/4}=U/2\pm \sqrt{U^2/4+1} \ .
\label{poles2}
\eeq
After transforming back to the time-dependent behavior, the dynamics
of the spin imbalance is characterized by only the two frequencies $z_3,z_4$:
\[
\langle\Psi_t|S^z_1-S^z_2|\Psi_t\rangle
=c_1\cos(z_3t)+c_2\cos(z_4t) 
\]
The constant term vanishes because the numerator in Eq. (\ref{spinimb3}) is zero for $z=z'=0$.
Moreover, the coefficients are
\[
c_1=-\frac{2}{\sqrt{U^2+4}(U-\sqrt{U^2+4})}\ ,\ \ \
c_2=\frac{2}{\sqrt{U^2+4}(U+\sqrt{U^2+4})}
\ .
\]
The spin imbalance as a function of time and the corresponding energy levels are plotted
in Fig. \ref{spinimbalance} for $U=2$ and two different values of $J/U$.
The coefficient of the higher frequency is substantially smaller than that
of the lower frequency. Moreover, with increasing $J/U$ the lower frequency
as well as the amplitude of the higher frequency increases.
This behavior is very similar to the experimental observation of the spin imbalance
by Trotzky et al. \cite{trotzky08}.

The return probability is calculated from the matrix element
\[
\langle\u,\d|(z-H'_0(z))^{-1}|\u,\d\rangle
=\frac{2zU-2z^2+1}{2z(zU-z^2+1)}
\]
which again has the poles $z_2,z_3,z_4$ of Eq. (\ref{poles2}) like the spin imbalance.
This leads to the time-dependent behavior
\[
\langle\Psi_0|\Psi_t\rangle 
=C_0+C_1e^{-iz_3t}+C_2e^{-iz_4t} \ ,
\]
where $C_0=1/2$ and
\[
C_1=\frac{1}{(U+\sqrt{U^2+4})\sqrt{U^2+4}} \ ,\ \ \ 
C_2=-\frac{1}{(U-\sqrt{U^2+4})\sqrt{U^2+4}}
\ .
\]
Then the time-dependent behavior of the return probability is characterized 
by three different frequencies:
\beq
P_t=C_0^2+C_1^2+C_2^2+2C_0C_1\cos(z_3t)+2C_0C_2\cos(z_4t)
+2C_1C_2\cos[(z_3-z_4)t]
\ .
\label{return2}
\eeq
The ratio of the coefficients $C_2$, $C_1$ is
\beq
\frac{C_2}{C_1} 
=\frac{\sqrt{1+4/U^2}+1}{\sqrt{1+4/U^2}-1} \sim U^{2}
\label{ratio}
\eeq
with the asymptotic behavior for $U\sim\infty$. Thus for
sufficiently large interaction the oscillating dynamics of $\langle\Psi_0|\Psi_t\rangle$
is dominated by the lower frequency $z_4=U/2- \sqrt{U^2/4+1}$, whereas for weaker
interaction two frequencies contribute with similar weight, namely 
$z_{3/4}=U/2\pm \sqrt{U^2/4+1}$.

{\it Falicov-Kimball limit}:
In the case of the asymmetric Falicov-Kimball limit the eigenvalues of 
Eq. (\ref{eigenvalues2}) are doubly degenerate for all values of $J_\d$.
This has some consequence for the dynamics. In particular, the spin imbalance 
of Eq. (\ref{2sh}) becomes
\[
\langle\u,\d|(z-H'_0(z))^{-1}(S^z_1-S^z_2)(z'-H'_0(z'))^{-1}|\u,\d\rangle 
=\frac{4(U-z')(U-z)}{(2Uz-2z^2+a)(2z'U-2z'^2+a)}
\]
and the transition matrix element reads 
\[
\langle\u,\d|(z-H'_0(z))^{-1}|\u,\d\rangle
=2\frac{U-z}{-2z^2+2zU+a} \ .
\]
Thus the expression of spin imbalance is the product of two transition matrix elements.
After transforming back to time we get
\[
\langle\Psi_0|\Psi_t\rangle
=C_1e^{-iz_3t}+C_2e^{-iz_4t}
\]
with $C_1=(z_3-U)/(z_3-z_4)$, $C_2=-(z_4-U)/(z_3-z_4)$, and
$z_{3/4}=U/2\pm\sqrt{U^2/4+J_\d^2}$. 
Then the spin imbalance is identical to the return probability:
\[
\langle\Psi_t|S^z_1-S^z_2|\Psi_t\rangle=|\langle\Psi_0|\Psi_t\rangle|^2\equiv P_t
\ ,
\]
where the oscillatory behavior is characterized by a single frequency:
\[
P_t=C_1^2+C_2^2+2C_1C_2\cos\left(\sqrt{U^2+4J_\d^2}t\right)
\ .
\]

\subsection{Two-site Bose-Hubbard model}

According to the Hartee approximation of the double-well potential \cite{milburn97},
the spectral properties change qualitatively when the number of bosons exceeds a
critical value $N_c\approx J/U$, where the regime with $N>N_c$ is characterized by 
``self trapping''. This is a regime in which the system stays in its initial state 
for arbitrarily long times.
For $N<N_c$, on the other hand, the regime is characterized by an oscillating behavior
with frequencies related to the tunneling rate. When the number of particles $N$
approaches the critical value $N_c$, the frequency of the oscillations goes down
to zero, indicating a real critical behavior. However, this might be an artifact of
the classical nonlinear equation obtained by the Hartree approximation. For a quantum
system on a finite-dimensional Hilbert space we expect no genuine critical behavior. 
Nevertheless, a crossover between two different regimes is possible, where in one regime
the time scale for escaping from the initial states can be very large and the escape 
is very unlikely. Such a behavior will be studied in the following.

\begin{figure}
\begin{center}
\includegraphics[width=7cm,height=7cm]{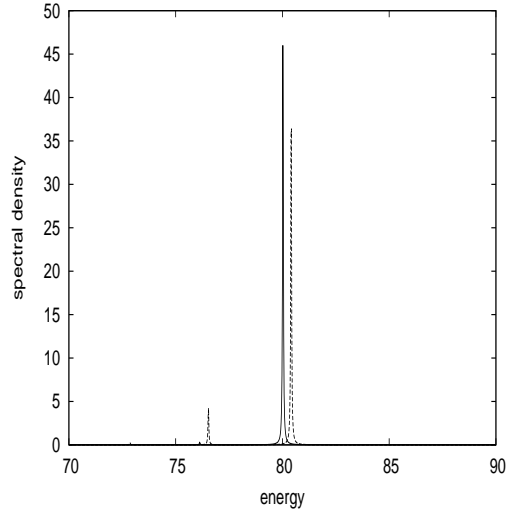} 
\caption{The spectral density of the BH model of $N=40$ bosons, $U=0.05$, $\epsilon=0.02$.
The height of the peaks corresponds to $|\langle E_j|\Psi_0\rangle|^2/\epsilon$ for 
the initial state $|\Psi_0\rangle$ (cf. text). For $J=0.05$ (full curve) and $J=0.2$ (dashed curve)
there is only one dominant energy level, out of 41 energy levels. This is a signature of ``self trapping''
due to strong interaction ($J/U=1$).
}
\label{plots6}
\end{center}
\end{figure}

\begin{figure}
\begin{center}
\includegraphics[width=7cm,height=7cm]{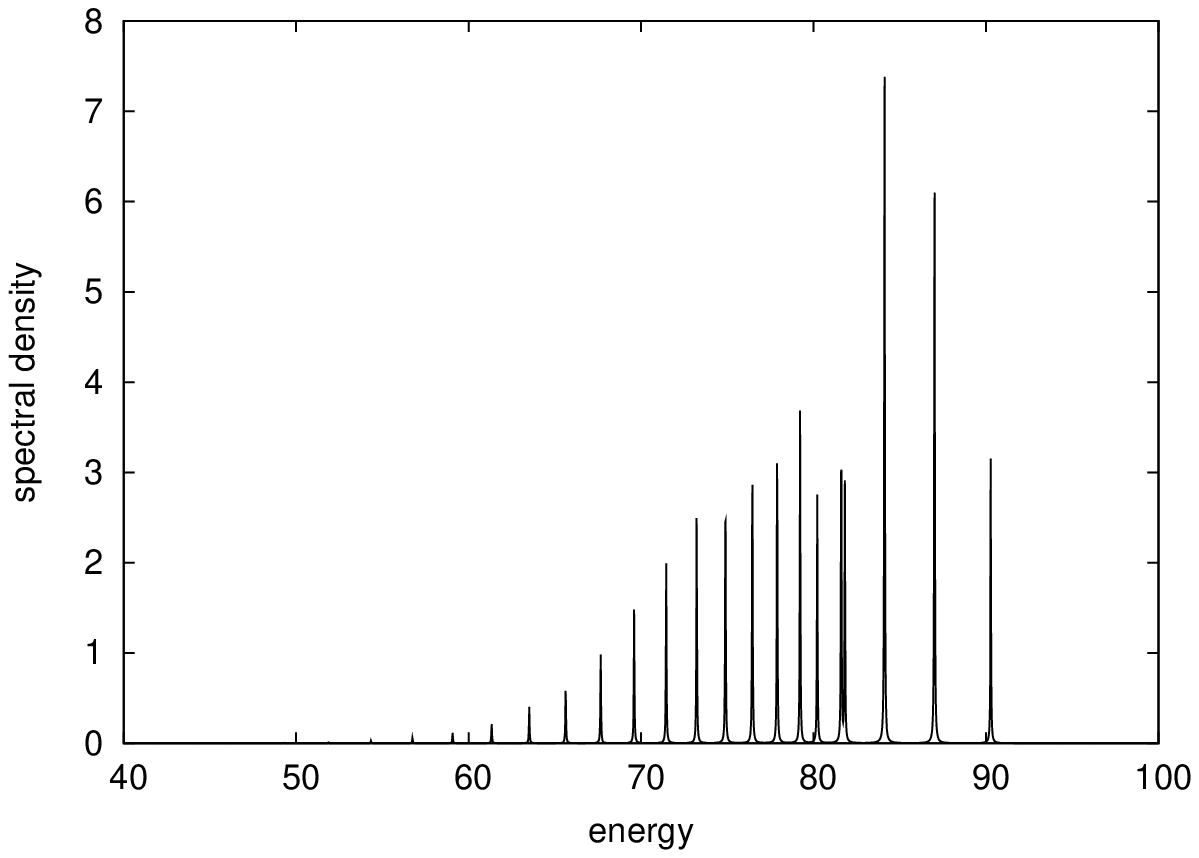} 
\includegraphics[width=7cm,height=7cm]{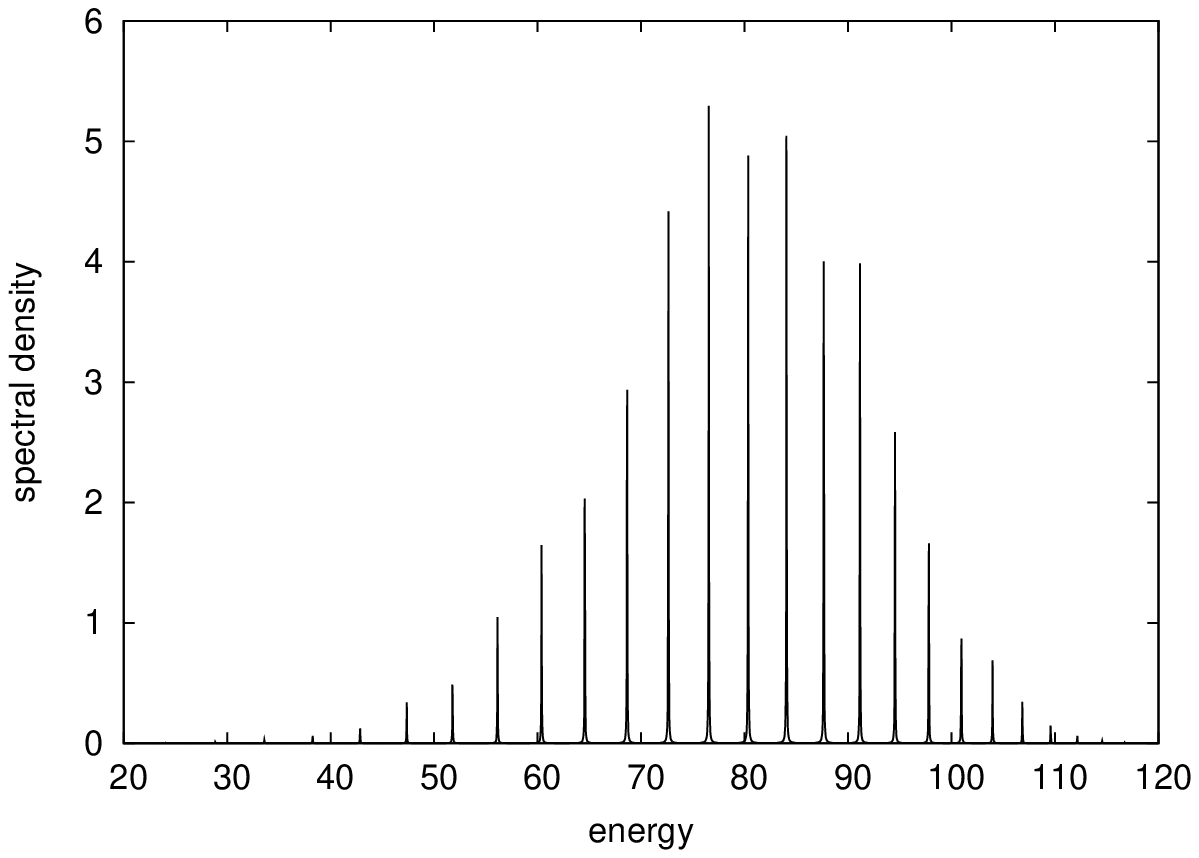} 
\caption{The spectral density of the BH model of $N=40$ bosons, $U=0.05$, $\epsilon=0.02$, 
$J=1$ (left panel) and $J=2$ (right panel). This clearly indicates that with increasing $J$ more
and more states can be reached dynamically with reasonable probability. 
Moreover, the spectral density is becoming more symmetric. However, even for $J/U=40$ only
about half of the 41 energy levels are visible in the plot.
}
\label{plots7}
\end{center}
\end{figure}

There are two types of quantities that determine the dynamical behavior on a 
finite-dimensional Hilbert space. These are the energy levels $E_k$ and the transition 
probabilities $|\langle\Psi_0|E_k\rangle|^2$ between the initial state and the eigenstates
of the Hamiltonian. 
Both quantities appear explicitly in the many-body spectral density of 
Eq. (\ref{lorenz}).
For the initial state $|N,0\rangle$ the many-body spectral density
is plotted for several parameter values in Figs. \ref{plots6}, \ref{plots7}.
This clearly indicates that the distribution of transition probabilities is
broad for weak interaction, referring to a complex oscillating dynamics,
and becomes narrower with increasing $U$.
For sufficiently large $U$ all transition probabilities are extremely small except
for one (cf. Fig. \ref{plots6}). This implies that the system cannot escape from 
its initial state. This strong-interaction behavior can be linked to the semi-classical 
self trapping of the Hartree approximation.

\begin{figure}
\begin{center}
\includegraphics[width=7cm,height=7cm]{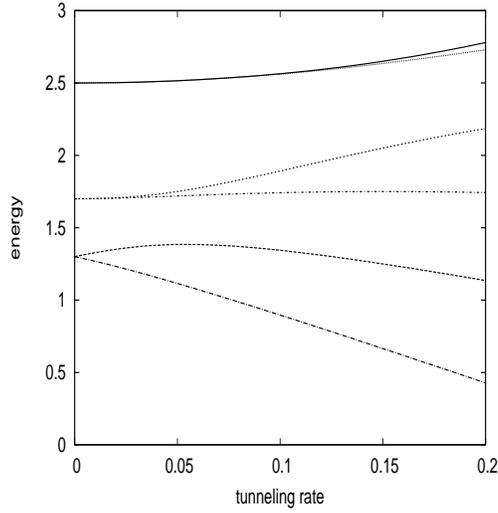} 
\caption{The six energy levels of the BH model of $N=5$ bosons, as a function of the tunneling rate $J$ at $U=0.1$.
At $J=0$ there are three doubly degenerate eigenvalues. The degeneracies are lifted by the tunneling rate $J$:
The higher the energy, the weaker the lifting of the degeneracies.
This behavior is very different from the non-interacting case $U=0$, where all energy levels behave linearly with $J$: 
$E=\pm J,\pm 3J, \pm 5J$. 
}
\label{plots5}
\end{center}
\end{figure}


\section{Conclusions}

The dynamics of finite FH and BH models is characterized by a discrete
set of energy levels and non-zero overlaps of the eigenfunctions with 
the initial state. By employing the recursive projection method we have 
calculated these quantities in terms of a many-body spectral function
for a double well. Moreover, the return probability
and the spin imbalance are obtained for the spin-1/2 FH model by the
same method. Depending on the details of the spin-dependent tunneling 
process, we have found a single-frequency dynamics (Falicov-Kimball limit)
and a two-frequency dynamics (symmetric FH model). In particular,
the spin imbalance of the symmetric FH model with tunneling rate $J$
reveals an oscillating behavior with the characteristic frequencies 
$U(1\pm\sqrt{1+16J^2/U^2})/2$, where the amplitude of the higher frequency
decreases strongly with increasing interaction $U$. This behavior agrees 
well with recent experimental observations \cite{trotzky08}.

For the double well with $N$ spinless bosons the dynamics indicates a 
crossover from a many-frequency dynamics for weak interaction to a 
stationary behavior at strong interaction. The latter can be related to 
the self-trapping behavior found in the Hartree approximation.

\begin{acknowledgments}
The author is grateful to Prof. A.F.R. de Toldeo Piza
for bringing Ref. \cite{feshbach67} to his attention.
This work was supported by the Aspen Center for Physics.
\end{acknowledgments}

\appendix

\section{Recurrence relation for the two-site Bose-Hubbard model}
\label{rpmbose}

For bosonic operators $b$ and $b^\dagger$ and particle-number states $|n\rangle$
we have 
\[
b|n\rangle=\sqrt{n}|n-1\rangle, \ \ b^\dagger|n\rangle=\sqrt{n+1}|n+1\rangle 
\ .
\]
Then a recursion step of the RPM the tunneling term $H_J$ changes the number of
particles in each well by one: 
\[
H_J|n_1,n_2\rangle
=-J(b_2^\dagger b_1+b_1^\dagger b_2)|n_1,n_2\rangle
=-J(\sqrt{(n_2+1)n_1}|n_1-1,n_2+1\rangle+\sqrt{n_2(n_1+1)}|n_1+1,n_2-1\rangle)
\ .
\]
Therefore, ${\cal H}_{2j}$ is two dimensional and spanned by 
$\{|n_1-j,n_2+j\rangle,|n_1+j,n_2-j\rangle\}$, provided $n_j-j\ge0$.
A special case is $n_1=N$, $n_2=0$. Then the projected spaces are one dimensional
and spanned by $|N-j,j\rangle$. The projected Hamiltonian gives
\[
P_{2j}J(b_2^\dagger b_1+b_1^\dagger b_2)|N-j,j\rangle
=Jb_2^\dagger b_1|N-j,j\rangle
=J\sqrt{(j+1)(N-j)}|N-j-1,j+1\rangle
\]
\[
=J\sqrt{(j+1)(N-j)}|N-(j+1),j+1\rangle
\]
and the recurrence relation reads
\[
\langle N-j,j|G_{2j}|N-j,j\rangle=\frac{1}{z-\langle N-j,j|H'_{2j}|N-j,j\rangle}
\ .
\]
The diagonal matrix elements of the effective Hamiltonians 
$\langle N-j,j|H'_{2j}|N-j,j\rangle$ are
\[
\langle N-j,j|H'_{2j}|N-j,j\rangle
=U[(N-j)^2+j^2]+J^2(j+1)(N-j)\langle N-j-1,j+1|G_{2j+2}|N-j-1,j+1\rangle
\]
and
\[
\langle 0,N|H'_{2N}|0,N\rangle=UN^2
\ .
\]
Using the notation $g_{N-j}\equiv\langle N-j,j|G_{2j}|N-j,j\rangle$, we
get from the recurrence relation
\[
g_{N-j}=\frac{1}{z-U[(N-j)^2+j^2]-J^2(j+1)(N-j)g_{N-j-1}}
\ .
\]
Finally, we can use the notation $k=N-j$ which implies $j=N-k$ and the recurrence relation
for $k=0,1,...,N$
\beq
g_k=\frac{1}{z-U[k^2+(N-k)^2]-J^2(N-k+1)kg_{k-1}}, \ \ \ g_0=\frac{1}{z-UN^2}
\eeq
and
\[
g_N=\langle N,0|G_0|N,0\rangle 
\ .
\]


\begin{thebibliography}{99}

\bibitem{foelling07}
S. F\"olling et al., Nature 448, 1029 (2007)

\bibitem{spielman07}
I. B. Spielman, W. D. Phillips, and J. V. Porto
Phys. Rev. Lett. {\bf 98}, 080404 (2007)

\bibitem{salger07}
T. Salger, C. Geckeler, S. Kling and M. Weitz,
Phys. Rev. Lett. {\bf 99}, 190405 (2007)

\bibitem{trotzky08}
S. Trotzky et al., 
Science 319, 295 (2008)

\bibitem{wuertz09}
P. W\"urtz et al., 
Phys. Rev. Lett. {\bf 103}, 080404 (2009)

\bibitem{shin06}
Y. Shin et al., 
Phys. Rev. Lett. 97, 030401 (2006);
G. B. Partridge et al., 
Phys.Rev.Lett. {\bf 97}, 190407 (2006);
N. Strohmaier et al., 
Phys. Rev. Lett. 99, 220601 (2007)

\bibitem{fermischool07}
{\sl Ultra-cold Fermi Gases}, Eds. M. Inguscio, W. Ketterle and C. Salomon,
IOS Press (Amsterdam 2007)

\bibitem{lewenstein07}
M. Lewenstein et al., Adv. Phys. {\bf 56}, 243


\bibitem{cohen}
C. Cohen-Tannoudji, J. Dupont-Roc, G. Grynberg, {\sl Atom-Photon Interactions},
John Wiley (New York 1992)

\bibitem{esteve08}
J. Est\'eve, C. Gross, A. Weller, S. Giovanazzi and M.K. Oberthaler,
Nature 455, 1216-1219 (30 October 2008) 

\bibitem{cirac07}
J.I. Cirac, in {\sl Ultra-cold Fermi Gases}, Eds. M. Inguscio, W. Ketterle and C. Salomon,
IOS Press (Amsterdam 2007)

\bibitem{duan03}
L.-M. Duan, E. Demler, M.D. Lukin, Phys. Rev. Lett. {\bf 91}, 090402 (2003)

\bibitem{rey07}
A.M. Rey et al., Phys. Rev. Lett. {\bf 99}, 140601 (2007)


\bibitem{milburn97}
G.J. Milburn et al., Phys. Rev. A {\bf 55}, 4318 (1997)

\bibitem{zoellner08}
S. Z\"ollner, H.-D. Meyer, P. Schmelcher, Phys. Rev. A {\bf 78}, 013621 (2008)

\bibitem{georges07}
W. Metzner and D. Vollhardt, Phys. Rev. Lett. {\bf 62}, 324 (1989);
A. Georges et al., Rev. Mod. Phys. {\bf 68}, 13 (1996);
A. Georges, in {\sl Ultra-cold Fermi Gases}, Eds. M. Inguscio, W. Ketterle and C. Salomon,
IOS Press (Amsterdam 2007)

\bibitem{feshbach67}
H. Feshbach, A. K. Kerman and R. H. Lemmer,
Annals of Physics {\bf 41}, 230 (1967)

\bibitem{ziegler}
K. Ziegler, Phys. Rev. A {\bf 68}, 053602 (2003); Phys. Rev. B {\bf 72}, 075120 (2005);
Phys. Rev. A {\bf 77}, 013623 (2008)

\bibitem{lanczos}
A.S. Householder, {\sl Theory of Matrices in Numerical Analysis},
Dover (New York 1974)

\bibitem{falicov69}
L.M. Falicov and J.C. Kimball, Phys. Rev. Lett. {\bf 22}, 997 (1969) ;
J.K. Freericks, Phys. Rev. B {\bf 48}, 3881 (1993);
P. Farkasovsky, Z. Phys. B {\bf 102}, 91 (1996); Eur. Phys. Lett. {\bf 84}, 37010 (2008);
J.K. Freericks and V. Zlati\'c, Rev. Mod. Phys. {\bf 75}, 1333 (2003)

\bibitem{fradkin}
E. Fradkin, {\sl Field Theories of Condensed Matter Systems}
Addison Wesley (1991)

\end{thebibliography}
\end{document}